\documentclass[iop,onecolumn,preprint,superscriptaddress,showpacs,nofootinbib]{revtex4}
\usepackage{color}
\usepackage{subeqnarray}
\usepackage{amsmath,amssymb}
\usepackage{color}
\usepackage[pdftex]{graphicx} 
\usepackage{hyperref}
\usepackage{graphicx}
\def\laq{\raise 0.4ex\hbox{$<$}\kern -0.8em\lower 0.62ex\hbox{$\sim$}}
\def\gaq{\raise 0.4ex\hbox{$>$}\kern -0.7em\lower 0.62ex\hbox{$\sim$}}
\newcommand{\be}{\begin{equation}}
\newcommand{\ee}{\end{equation}}
\newcommand{\bea}{\begin{eqnarray}}
\newcommand{\eea}{\end{eqnarray}}

\begin{document}
\title{Three mode interaction noise in laser interferometer gravitational wave detectors }
\author{Li Ju}
\email{li.ju@uwa.edu.au}
\affiliation{School of Physics, University of Western Australia, WA 6009, Australia}
\author{Chunnong Zhao}
\email{chunnong.zhao@uwa.edu.au}
\affiliation{School of Physics, University of Western Australia, WA 6009, Australia}
\author{Yiqiu Ma}
\email{myqphy@gmail.com}
\affiliation{School of Physics, University of Western Australia, WA 6009, Australia}
\author{David Blair}
\affiliation{School of Physics, University of Western Australia, WA 6009, Australia}
\author{Stefan.L.Danilishin}
\affiliation{School of Physics, University of Western Australia, WA 6009, Australia}
\author{Slawek Gras}
\affiliation{Kavli Institute for Astrophysics and Space Research, Massachusetts Institute of Technology, 02139, USA}
\begin{abstract}
Triply resonant three mode interactions in long optical cavities have been shown to lead to enhanced scattering of carrier light by the ultrasonic acoustic modes of the test mass mirrors. At high optical power, this can lead to parametric instability (parametric gain $R>1$) for a few acoustic modes with strong spectral and spatial overlap. Numerous $\sim10^3$ acoustic modes of the test masses are predicted to have $R>10^{-2}$.  Experimental studies have shown that such modes also strongly scatter the carrier light, enabling very sensitive readout of the acoustic modes. The 3-mode scattering from the thermal fluctuation of large population of ultrasonic modes would causes random changes in occupation number of the carrier light and cavity transverse optical modes. Because the thermal fluctuation time scale (set by the acoustic mode relaxation times) is typically a few seconds, the noise spectrum from thermally induced photon number fluctuations is strongly peaked at low frequency.  The noise level depends on the acoustic mode structure and acoustic losses of the test masses, the transverse optical mode spectrum of the optical cavities and on the test mass temperature. We theoretically investigate the possible effect of this noise and show that in advanced detectors under construction three mode interaction noise is below the standard quantum limit, but could set limits on future low frequency detectors that aim to exceed the free mass standard quantum limit.
\end{abstract}

\maketitle
\section{Introduction}
At low frequencies gravitational wave detectors such as Advanced LIGO~\cite{GHarry}, advanced Virgo~\cite{Accadia} and KAGRA~\cite{Somiya} are expected to be limited by quantum radiation pressure fluctuations due to photon number fluctuations in their laser beams. The fluctuations in radiation pressure forces act on the test masses and create a noise spectrum which increases at low frequency. Much effort is underway to develop quantum measurement schemes that could overcome this fundamental noise source~\cite{Buonanno,Zach,Kimble,Stefan}.  Third generation detectors are proposed that would increase low frequency sensitivity down towards 1Hz~\cite{Punturoet}.  Here we discuss a mechanism by which the thermal fluctuations of acoustic modes lead to additional fluctuations through three mode interactions.

Three mode interactions in laser interferometer gravitational wave detectors were first considered by Braginsky et al in 2001~\cite{Braginsky2001,Braginsky2002}. He recognized that the scattering described by the diagrams in Figure 1 was likely to occur in long optical cavities in which the free spectral range was comparable to the internal acoustic mode frequencies of the test masses. Since Braginsky's prediction, there have been many further detailed studies that confirmed the model~\cite{Zhao2005,Strigin2007,Evans2010,Gras2010} and a succession of experimental studies have verified the theory in detail~\cite{Zhao2011,Sunil2013,Blair2013,Xu2013}.

The parametric gain for three mode interactions (derived by ~\cite{Braginsky2001}) describes the fraction of ring down power of a test mass internal mode that is returned to the test mass by radiation pressure feedback. Positive gain, characterised by the parametric gain $R$, causes heating of a test mass mode and if $R> 1$ the feedback leads to instability. Negative gain $(R<0)$ corresponds to negative feedback, and causes the cooling of acoustic modes.

The feedback occurs as follows. An acoustic mode (frequency $\omega_m$) of the mirror scatters the carrier light (frequency $\omega_0$). This process  would create two sidebands of frequency $\omega_0+\omega_m$ and $\omega_0-\omega_m$. However when the optical linewidths are narrow compared with $\omega_m$,  the intrinsically asymmetric transverse mode structure of an optical cavity means that only one sideband is likely to be supported. The sideband light beats with the carrier light to create a fluctuating radiation pressure force on the mirror. If the lower sideband is supported, the scattering depletes the carrier, and the beat frequency is in phase with the acoustic mode, giving rise to positive feedback. If the upper sideband frequency is coincident with a transverse mode, the beat frequency is out of phase with the acoustic mode and energy is extracted from this mode. From a quantum mechanical viewpoint, the two possible three mode processes (normally called Stokes and anti-Stokes processes) can be illustrated by the Feynman diagrams shown in figure 1.

\begin{figure}[!t]
\begin{center}
\includegraphics[width=0.8\textwidth]{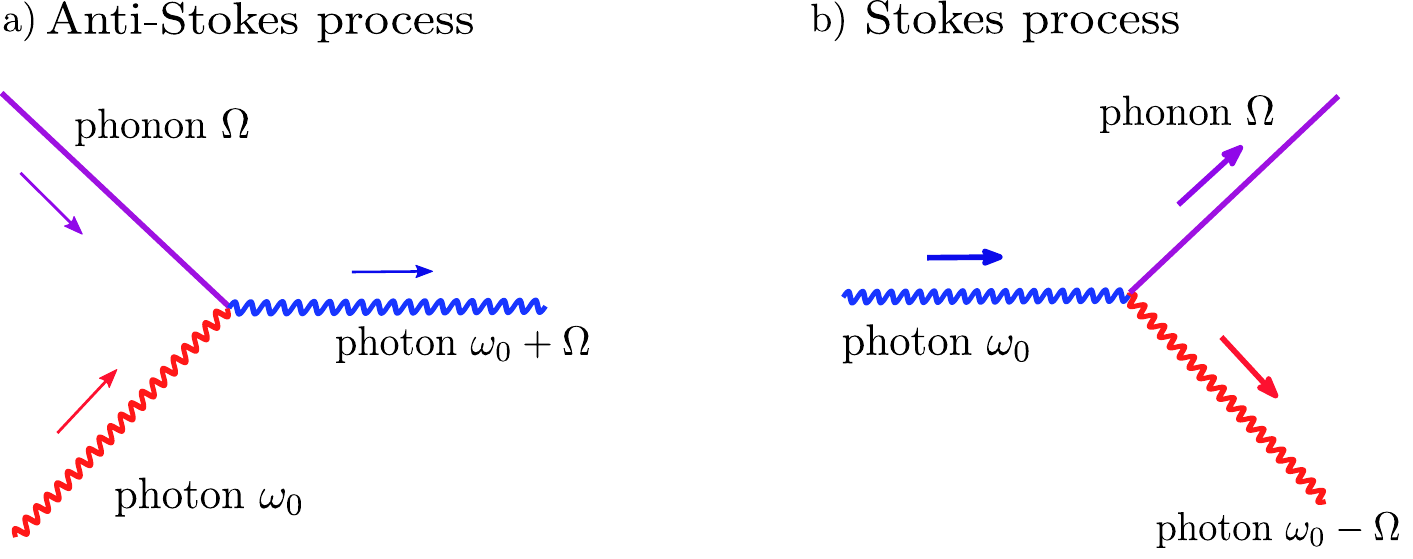}
\caption{Feynman diagram for three mode interactions showing both heating and cooling processes. The left
hand diagram shows a cooling process in which an acoustic phonon is absorbed to create a transverse mode
photon of frequency $\omega_0+\omega_m$. The right hand diagram shows a positive feedback interaction in which the
photon is down converted to $\omega_0-\omega_m$ and a phonon which is emitted into the test mass. }
\label{fig1}
\end{center}
\end{figure}

In this paper we are interested in the effect of both types
of feedback processes. Our concern is not about parametric
instability, which we assume is always controlled by various
techniques~\cite{Juli2009,Miller2011,Gras2009,Gras2013,Evans2008}.
The interaction enhances the scattering in the positive feedback domain and suppresses the scattering in the negative feedback domain. For example, for $R=0.1$, the corrections for feedback will be about 10 percent, but if $R\rightarrow 1$, the feedback can cause large amplification.

In an advanced gravitational wave detector such as Advanced LIGO, Advanced Virgo~\cite{GHarry,Accadia}, the dense spectrum of acoustic modes scatters photons out of the main laser beam through Stokes and anti-Stokes processes associated with each acoustic modes. Both Stokes and anti-Stokes processes contribute to fluctuations of the photon number in the interferometer cavities. Using results from detailed modelling of parametric interactions in typical advanced interferometer cavities~\cite{Gras2010}, we will show that out of 5500 acoustic mode in the frequency range 5kHz to 150kHz, there are $\sim10$ acoustic modes with $R>1$ and 70 acoustic modes with $R >10^{-1}$.  These modes are all independent and will simultaneously contribute random light scattering of the carrier~\cite{Juli2006} as the thermally excited acoustic mode amplitude fluctuates.

We will show that the total number of modes is roughly independent of the optical cavity configuration. We assume that all acoustic modes are thermally excited and in thermal equilibrium with the reservoir, and hence have mean energy $k_BT$.  The coupling of each mode to the thermal reservoir is determined by the acoustic quality factor $Q_i$ of each mode. This also defines the rate of fluctuation of the mode amplitude. We investigate whether the combined effect of this scattering leads to non-negligible photon number fluctuations.

In this paper we estimate the amplitude of this effect for a typical advanced interferometer. We will show that this noise does not currently threaten the sensitivity of planned detectors, because its $1/f$ spectrum that is generally below the quantum noise.

First we will give an order of magnitude estimation of the magnitude of this noise source. Then we  will derive the spectrum of this force noise from the first principles. Next present modelling data for the number of acoustic modes as a function of their parametric gain. While this quantity in principle depends on the precise mirror radii of curvature, we find that in practice the total number of modes is remarkably constant over a range of mirror radii of curvature. Finally we will estimate the total light scattering from a distribution of acoustic modes. We will assume that all modes have a quality factor typical of recent observations, and show that the results do not depend very strongly on the actual  acoustic mode quality factors because increased losses reduce the parametric gain, and hence the total scattering, while simultaneously increasing the bandwidth of the fluctuations. We will also show that the noise level due to three mode interactions increases linearly with optical power level, in contrast to quantum noise that increases as the square root of power. This means that this noise source would ultimately exceed the quantum noise level at extremely high power levels.

\section{Theoretical modelling}
\subsection{Simple estimation}
It is possible to make a simple estimate of the magnitude of three mode interaction noise based on previous modelling of parametric instability~\cite{Gras2013}. We start from the fact that a parametric gain of unity requires the optical power loss from the cavity to equal the power dissipation of an acoustic mode. If the mode energy is $k_BT$, then the energy loss is $2\pi k_BT/Q$ per cycle, where $T$ is the test mass temperature and $Q$ is the acoustic mode quality factor. In general the energy loss from the optical system for a mode with parametric gain R is $2\pi Rk_BT/Q$ per cycle. The energy lost from the cavity pump mode per cycle is larger than this by a factor of pumping mode frequency over acoustic mode frequency $\omega_0/\omega_m$ since the acoustic power is derived from the frequency difference of the two optical modes.  (We assume that the transverse optical mode is rapidly lost from the optical system and hence does not contribute to the radiation pressure).

For the $i_{th}$ acoustic mode of gain $R_i$, the total power loss (energy loss per second) from the cavity would be
\be
\Delta P_{\text{i-total}}=\frac{2\pi R_ik_BT}{Q_i}\frac{\omega_0}{\omega_{mi}}
\frac{\omega_{mi}}{2\pi}=\frac{\omega_0k_BTR_i}{Q_i}.
\ee
Making the approximation that the total power loss is uniform within the acoustic mode linewidth $2\gamma_{mi}=\omega_{mi}/Q_i$, then the spectral density of the power loss is,
\be
\Delta P_i(\omega_{mi})=\frac{\omega_0k_BTR_i}{\omega_{mi}},
\ee
The sum of all the interaction induced power loss is then:
\be
\Delta P=\omega_0k_BT\sqrt{\sum_i\left(\frac{R_i}
{\omega_{mi}}\right)^2},
\ee
This power loss can be equated to the power fluctuation or equivalently to the photon number fluctuation in 1 second, which we use for comparison with quantum power fluctuations $\Delta P_q$. With parameters $T\sim300$K, $k_B\sim10^{-23}$ and $\omega_{mi}\sim10^6$, and with the fact that there are many acoustic modes with non-negligible parametric gain, the magnitude of photon fluctuation $\Delta P_q/\hbar\omega_0$ can be seen to be $\sim10^{8}$. However the typical acoustic mode relaxation time is several seconds. In addition, modes with high values of $|R|$ change the mode temperatures by the parametric amplification process. A spectral noise estimate is necessary to determine whether this noise has any impact on Advanced GW detectors.

\subsection{Hamiltonian and the equation of motions}
Now we theoretically analyze the three-mode interaction noise due to the thermal motion of one mechanical vibration mode. Consider a surface-vibration-modulated high-order mode resonant inside the cavity (in the form of the Stokes sidebands for a positive parametric gain and anti-Stokes sidebands for negative parametric gain), then a multi-mode Hamiltonian can be given by:

\be\label{Hamiltonian}
\hat{H}=\hbar\omega_0\hat a^{\dagger}\hat a+\hbar\omega_1\hat b^{\dagger}\hat b+\hbar G_0\hat x(\hat a^{\dagger}\hat b+h.c)+\hbar G_c(\hat a^{\dagger}\hat a+\hat b^{\dagger}\hat b)\hat X+H_m+H_{ext}.
\ee
 The $\hat a$, $\hat b$, $\hat x$, $\hat X$ denote the pumping mode with resonant frequency $\omega_0$, high-order mode
 with resonant frequency $\omega_1$, mechanical displacement of the internal mechanical mode and center of mass motion, respectively.
 The optomechanical three-mode interaction strength $G_0$ is given by $\sqrt{\Lambda \omega_0\omega_1/L^2}$ where $\Lambda$ is the mode overlap factor and $L$ is the cavity length. The three mode interaction noise induced by a three-mode parametric coupling between the thermal-fluctuating $\hat x$ and $\hat a,\hat b$ fields is given in the third term of Eq.\ref{Hamiltonian}. Only $\hat X$ is coupled to gravitational waves. Radiation pressure forces given by the fourth term come from the coupling of the optical fields to the center of mass motion $\hat X$ with strength $G_c=\omega_0/L$. Terms $H_m$ and $H_{ext}$ describe the free mechanical degrees of freedom and the coupling between our system and the external thermal/optical bath, respectively.

From the Hamiltonian, we can derive the equations of motion
in the rotating frame with the frequency $\omega_0$:
\begin{subequations}\label{eom}
\begin{align}
&M\ddot{\hat X}=-\hbar G_{c}(\hat{a}^{\dagger}\hat a+\hat b^{\dagger}\hat b)\label{eomc}\\
&m(\ddot{\hat x}+\gamma_m\dot{\hat x}+\omega_m^2\hat x)=-\hbar G_i(\hat b \hat a^{\dagger}+h.c)+\xi^{th}\label{eomd}\\
&\dot{\hat a}+\gamma_0\hat a=-iG_{c}\hat a\hat X-iG_0\hat x\hat b+\sqrt{2\gamma}\hat a_{in}\label{eoma}\\
&\dot{\hat b}+(-i\Delta_1+\gamma_1)\hat b=-iG_0\hat x\hat a-iG_{c}\hat b\hat X+\sqrt{2\gamma_1}\hat b_{in}\label{eomb},
\end{align}
\end{subequations}
in which $M$ is the mass of the whole mirror and $m$ is the effective mass of the acoustic mode; $\gamma_m$, $\omega_m$ and $\xi^{th}$are the mechanical bandwidth, resonant frequency and the stochastic thermal force which drives the internal modes. The $\gamma_0$ and $\gamma_1$ are the bandwidth of pumping mode and the high-order mode. The detuning of
the pumping beam with respect to the resonance of the high- order mode is given by $\Delta_1=\omega_0-\omega_1$. The terms $\hat a_{in}$
and $\hat b_{in}$ are the vacuum fluctuations of the eletromagnetic field injected into the fundamental mode and high-order mode channels respectively.

Since the $\hat a$ field is pumped by a strong laser, we can perturbatively solve the above problem. Firstly, taking the
equilibrium position of the mechanical mode as the zero reference point the steady amplitude of $\hat{a}$ and $\hat{b}$ fields are given by:
\begin{subequations}\label{meanfield}
\begin{align}
&\bar{b}=-iG_0\bar{a}\bar{x}_m=0\\
&\bar{a}=\sqrt{2/\gamma_0}\bar{a}_{in}=\sqrt{2I_0/\gamma_0\hbar\omega_0},
\end{align}
\end{subequations}
with $I_0$ be the power of the pumping beam.
Then the optical fields can be expanded as:
$\hat a=\bar{a}+\hat a^{(1)}+\hat a^{(2)}+...$ and
$\hat b=\bar{b}+\hat b^{(1)}+\hat b^{(2)}+...$. From
Eq.~\eqref{meanfield}, the zeroth-order of $\hat b$  does not exist. The radiation pressure force acting on the center of mass degree of freedom can be perturbatively expanded to second order as:
\be\label{rad}
\begin{split}
M\ddot{\hat X}=-\hbar G_c[|\bar a|^2+\bar{a}(\hat a^{(1)\dagger}+\hat a^{(1)})+\hat a^{(1)\dagger}\hat a^{(1)}+\bar{a}(\hat a^{(2)\dagger}+\hat a^{(2)})
+\hat b^{(1)\dagger}\hat b^{(1)+h.c}].
\end{split}
\ee
The last two terms on the right hand side of Eq.~\ref{rad} contribute to the three-mode-interaction noise.
 The first one is the stochastic loss of the pumping field $\hat a$ due to the thermal motion of the internal mechanical mode $x$. This comes from the second term on the right hand side of Eq.~\eqref{eoma}. Since the $\hat x$ and $\hat b$ are both contributed by thermal fluctuations that can be considered as a perturbation to the mean field value, this term is a \emph{2nd-order perturbative} quantity. The last term in Eq.~\ref{rad} is the fluctuation of the high-order optical mode $\hat b$ due to the three mode interaction. It comes from the first term on the right hand side of Eq.~\eqref{eomb}. Because of the strong pumping, $\bar{a}$ is of zeroth order. Hence this term is a \emph{1st-order perturbative} quantity. Note that, when $\hat b^{(1)}$ contributes to the radiation pressure noise ( acting on the center of mass of the mirror ) the radiation pressure force is again a \emph{2nd-order perturbative} quantity because there is no zeroth order $\hat b-$field as we can see from Eq.~\eqref{rad}.

Although the 3-mode interaction noise is a 2nd order effect,
it is amplified due to the parametric gain. The parametric amplification process happens at the 1st order level, as discussed below. The set of the 1st order linearized equations of motion are as follows:
\begin{subequations}\label{eom1}
\begin{align}
&\dot{\hat a}^{(1)}+\gamma_0\hat a^{(1)}=-iG_{c}\bar a\hat X+\sqrt{2\gamma_0}\hat a_{in}^{(1)}\label{eoma1}\\
&\dot{\hat b}^{(1)}+(-i\Delta_1+\gamma_1)\hat b^{(1)}=-iG_0\hat x\bar a+\sqrt{2\gamma_1}\hat b_{in}^{(1)}\label{eomb1}\\
&M\ddot{\hat X}^{(1)}=-\hbar G_{c}\bar{a}(\hat a^{(1)\dagger}+\hat a^{(1)})\label{eomc1}\\
&m(\ddot{\hat x}^{(1)}+\gamma_m\dot{\hat x}^{(1)}+\omega_m^2\hat x^{(1)})=-\hbar G_i\bar{a}(\hat b^{(1)} +h.c)+\xi^{th(1)}\label{eomd1}.
\end{align}
\end{subequations}
The $\hat b^{(1)}$ field contains the displacement signal of the internal mode as shown in Eq.~\eqref{eomb1}, which will be fed back to the three mode interaction term in Eq.~\eqref{eomd1}. This feedback process will modify the dynamics of the internal mode, create a shift of mechanical resonant frequency (usually negligible compared to $\omega_m$) and an optical damping term $-m\gamma_{opt}\dot{\hat x}^{(1)}$ which can be positive or negative, corresponding to parametric heating or cooling. The parametric gain is defined as $R=-\gamma_{opt}/\gamma_m$~\cite{Haixing2009}.

From the linearized Eq.\eqref{eomb1}, we can derive the fluctuation field $\hat b^{(1)}$ and $\hat a^{(2)}$ (proportional to $\hat b^{(1)}\hat x$), which is the source of the three-mode interaction noise. From these two terms, we obtain the spectrum of the three-mode interaction noise. In the following section, we will discuss both terms.

\subsection{Noise contributed by the thermally induced fluctuations of high order mode}
The thermally induced fluctuations of the high order transverse mode $\hat b^{(1)}$ field is determined by Eq.\eqref{eomb1}. We choose a rotating frame at $\omega_m$ to describe $\bar b^{(1)}$ field and neglect the non-interesting quantum fluctuation part $\hat b^{(1)}_{in}$ which is not relevant to this calculation and is negligible. Then Eq.\eqref{eomb1}can be rewritten as:
\be
\dot{\hat b}^{(1)}=-(\gamma_1-i\Delta)\hat b^{(1)}-iG_0\hat x\bar ae^{i\omega_mt}.
\ee
In this case we define $\Delta=\Delta_1-\omega_m$,
and solve it in the time domain:
\be
\hat b(t)=-iG_0\bar{a}\int^{t}_{-\infty}e^{-(i\Delta+\gamma_1)(t-t^{\prime})}
e^{i\omega_mt^{\prime}}\hat x(t^{\prime})dt^{\prime}.
\ee
We can use the adiabatic approximation to move $e^{i\omega_m t}x(t)$ out of the integral since it is a slowly varying term. Also note that the frequency band of $e^{i\omega_m t}x(t)$ is of order $\gamma_m\ll\gamma_1$. This means that if $t^{\prime}$ is deviated from $t$ a little bit, then the contribution is already very small due to its rapid decay. Therefore the main contribution comes from $t^{\prime}\approx t$, that is:
\be\label{b(t)}
\hat b(t)\approx-iG_0\bar{a}\hat x(t)e^{i\omega_mt}\int^{t}_{-\infty}e^{-(i\Delta+\gamma_1)(t-t^{\prime})}dt^{\prime}.
=-\frac{-iG_0\bar{a}\hat x(t)e^{i\omega_mt}}{i\Delta+\gamma_1}
\ee
Then we  substitute this time domain solution into the last term of Eq.\ref{rad} to obtain:
\be
F_b(t)\propto(\hat b(t)^{\dagger}\hat b(t)+h.c)=-2\hbar G_0^2 G_c\left[\frac{\bar{a}^2}{\Delta^2+\gamma_1^2}
\right]\hat x(t)^2.
\ee
The correlation function of this force noise $\text{Cov}_{FF}(t,t^{\prime})$ is:
\be\label{covFF}
\text{Cov}_{FF}(t,t^{\prime})=\frac{1}{2}\langle F_b(t)F_b(t^{\prime})+F_b(t^{\prime})F_b(t)\rangle
=4\hbar^2 G_0^4 G_c^2\left[\frac{\bar{a}^4}{(\Delta^2+\gamma_1^2)^2}
\right]\langle \hat x(t)^2\hat x(t^{\prime})^2\rangle.
\ee
The two-point correlation function for $\hat x^2$ has been
given in the \emph{Appendix}. What we need to notice here is
that the mechanical response function is changed due to
the optomechanical feedback discussed in the last section. Now, the optomechanically modified mechanical response function (neglecting the negligible shift of mechanical resonant frequency) is:
$\chi_{\text{eff}}(\omega)=m(\omega_m^2-\omega^2+i(1-R)\gamma_m\omega)$.
Then the two-point correlation function of $x$ in the time domain becomes:
\be
S_{xx}(\tau)=\frac{2k_BT}{m\omega_m^2(1-R)}e^{-(1-R)\gamma_m\tau/2}
\cos{\omega_m\tau}.
\ee
Since the $x$ process is well approximated as a Gaussian process, therefore by Wick's theorem, we have:
\be\label{x2}
\langle \hat x(t)^2\hat x(t+\tau)^2\rangle=
\frac{4k_B^2T^2}{m^2\omega_m^4(1-R)^2}(1+e^{-(1-R)\gamma_m\tau/2}).
\ee
Taking a cosine transformation to the frequency domain ~\cite{Kip} and substituting it into Eq.~\eqref{covFF}, the spectrum of the force noise $F_b$ is given by:
\be\label{Sb}
S^{\text{3MIb}}_{\text{FF}}=\left(\frac{16G_c^2}{(\Delta^2+\gamma_1^2)^2}\right)
\left(\frac{k_B^2T^2}{\omega_m^2(1-R)}\right)\left(\frac{\gamma_1^2\gamma_m^3R^2}
{\Omega^2+(1-R)^2\gamma_m^2}\right).
\ee
Now we make use of the fact that:
$R=\hbar G_c^2\bar{a}^2/(m\omega_m\gamma_1\gamma_m)$.
In the case $\Delta\ll\gamma_1$ and substituting $Q=\omega_m/\gamma_m$, the the three mode interaction noise spectral density becomes:
\be
S^{\text{3MIb}}_{\text{FF}}=\frac{16\omega_0^2}{\gamma_1^2L^2}\left(\frac{k_B^2T^2}
{Q^2(1-R)}\right)\left(\frac{\gamma_mR^2}{\Omega^2+(1-R)^2\gamma_m^2}\right).
\ee
where $G_c$ has been replaced by $\omega_0/L$.

\subsection{Noise contributed by the thermally induced fluctuations of the pumping field}
We now repeat a similar analysis for the noise induced through fluctuations in the pumping field. The situation is different because
the high order mode force arises from the beating of two first order terms, while the pumping field fluctuation are caused by a beat between
the zeroth order pumping field and the second order fluctuating field. However, we will see that this leads to a similar result.

The 2nd-order perturbative equation of motion for the pumping $\hat a$ field is:
\be\label{a2}
\dot{\hat a}^{(2)}+\gamma_0\hat a^{(2)}=-iG_c(\bar{a}\hat X^{(2)}+\hat a^{(1)}\hat X^{(1)})-iG_0\hat x\hat b^{(1)}.
\ee
The three-mode interaction contributes to the last term on the right hand side of the above equation. Its corresponding part in $\hat a^{(2)}$ is defined to be $\hat a^{(2)}_{\text{th}}$.
The $\hat b^{(1)}(t)$ field is given by Eq.\eqref{b(t)} in the rotating frame with frequency $\omega_m$. By turning back to the non-rotating frame (by adding a $e^{i\omega_m t}$factor), substituting into $-iG_0\hat x\hat b^{(1)}$ in Eq.\eqref{a2}, and using the fact that $\gamma_m\ll\gamma_1$, we obtain:
\be\label{ath2}
\hat a^{(2)}_{th}(t)=-\frac{G_0^2\bar{a}}{i\Delta+\gamma_1}
\int^{t}_{-\infty}dt^{\prime}e^{-\gamma_0(t-t^{\prime})}
x^2(t^{\prime})\sim-\frac{G_0^2\bar{a}}{i\Delta+\gamma_1}
\frac{x^2(t)}{\gamma_0}.
\ee

The three mode interaction induced thermal radiation pressure force related to this $\hat a^{(2)}$ field is given by:
\be
F_{a}(t)=\hbar G_c\bar{a}(\hat a_{th}^{(2)\dagger}+\hat a_{th}^{(2)}).
\ee
Inserting Eq.\eqref{ath2}, and using the correlation function of $x(t)^2$ shown in Eq.\eqref{x2}, we can obtain the three mode interaction noise due to $\hat a^{(2)}$ as:
\be
S^{\text{3MIa}}_{\text{FF}}=\left(\frac{16G_c^2}{(\Delta^2+\gamma_1^2)^2}\right)
\left(\frac{k_B^2T^2}{\omega_m^2(1-R)}\right)\left(\frac{\gamma_m^3\gamma_1^4R^2}
{\gamma_0^2[\Omega^2+(1-R)^2\gamma_m^2]}\right).
\ee
In the case of $\Delta\ll\gamma_1$, substituting $Q=\omega_m/\gamma_m$, we obtain:
\be\label{Sa}
S^{\text{3MIa}}_{\text{FF}}=\left(\frac{16G_c^2}{\gamma_0^2}\right)
\left(\frac{k_B^2T^2}{Q^2(1-R)}\right)\left(\frac{\gamma_mR^2}
{\Omega^2+(1-R)^2\gamma_m^2}\right).
\ee
We see that in spite of the slightly different physics, Eq.~\ref{Sa} is identical to Eq.~\ref{Sb}
except that $\gamma_1$ is replaced by $\gamma_0$.

\subsection{Coherent cancelation}
In the above sections, we have derived the spectrum of three mode noise that comes from the $\hat b^{(1)}$ and $\hat a^{(2)}$ terms that represent the fluctuations of the higher order mode field and the pumping field due to three-mode interaction. In reality, these two noises are not independent of each other but strongly correlated. This point can be easily seen from the fact that in the three-mode parametric interaction process shown in Fig.~\ref{fig1}, the annihilation (creation) of a pumping field $\hat a$ photon will be accompanied with the creation (annihilation) of a high-order mode photon $\hat b$. Therefore, an increase of the radiation pressure force contributed by high-order mode will be accompanied with a decrease of the radiation pressure force contributed by the pumping mode. Therefore, the true radiation pressure force noise spectrum should be:
\be
\begin{split}\label{totalnoise}
S^{3MI}_{\text{FF}}=&\langle[F_a(t)- F_b(t)][ F_a(t^{\prime})- F_b(t^{\prime})]\rangle\\
=&\langle F_a(t)F_a(t^{\prime})\rangle+ \langle F_b(t)F_b(t^{\prime})\rangle-\langle F_a(t)F_b(t^{\prime})\rangle-\langle F_b(t)F_a(t^{\prime})]\rangle.
\end{split}
\ee
The first two terms in Eq.~\eqref{totalnoise} were calculated
in the above two subsections. The calculation of the last two terms is straightforward, following the same method, we obtain:
\be
\langle F_a(t)F_b(t^{\prime})\rangle=\langle F_b(t)F_a(t^{\prime})=\left(\frac{16G_c^2}{\gamma_0\gamma_1}\right)
\left(\frac{k_B^2T^2}{Q^2(1-R)}\right)\left(\frac{\gamma_m R^2}
{\Omega^2+(1-R)^2\gamma_m^2}\right).
\ee
The final result for the total noise is then:
\be\label{S3MIfinal}
S^{3MI}_{\text{FF}}=\left(\frac{16G_c^2k_B^2T^2}{Q^2(1-R)}\right)\left(\frac{\gamma_m R^2}{\Omega^2+(1-R)^2\gamma_m^2}\right)\left(\frac{\gamma_0-\gamma_1}
{\gamma_0\gamma_1}\right)^2.
\ee
Eq.~\ref{S3MIfinal} shows that if all higher order modes had the same linewidth as
the pumping mode, this noise would fall to zero.
It is useful to compare this result with the radiation pressure noise due to the
quantum vacuum fluctuations of electromagnetic field in the
low frequency region where $\Omega\sim(1-R)\gamma_m$. It is realistic to
assume that $\gamma_1$ is several times larger than $\gamma_0$ since higher order modes
usually have higher losses due to their larger spot size. This leads to an upper limit for $S^{3MI}_{\text{FF}}$, given by:
\be\label{ratio}
\frac{S^{3MI}_{\text{FF}}}{S^{\text{rad}}_{\text{FF}}}
\sim\left(\frac{ck_B^2T^2\omega_0}{L\hbar Q^2P_c
\gamma_1\gamma_m}\right)\left(\frac{R^2}{(1-R)^3}\right).
\ee
Here $P_c$ is the intra-cavity pumping power.
In the following section, we will use Eq.\eqref{S3MIfinal} to calculate the noise from a large number of acoustic modes.

For completeness, we discuss the validity of the perturbation method used here.
It is clear from Eq.\ref{S3MIfinal}, when $R\rightarrow1$, the
spectrum goes to infinity, which indicates that the perturbation method is no longer valid. This is due to the fact that
the optomechanical interaction heats or cools the mechanical degrees of freedom. Solving this problem fully requires simulation of the non-linear dynamics of the optomechanical system ~\cite{Xu2013}.
However, if the root mean square motion is small compared with the cavity's linear dynamical range, i.e:
\be\label{criteria}
\sqrt{\overline{x^2}}<\lambda/F,
\ee
($\sqrt{\overline{x^2}}$ is the root mean squre of mirror displacement due to the thermally excited acoustic mode and $F$ is the cavity finesse),
then the system will be within the linear dynamical region and the perturbation method will be valid.
For an advLIGO type detector, we have $\lambda/F\sim 10^{-9}$m. Since $\sqrt{\overline{x^2}}\sim 3\times 10^{-11}$m$^2$, even for $1-R\sim10^{-8}$ the system is well within the linear dynamical region. Thus we conclude that the perturbation method is valid in our discussion.

\section{Simulation results and discussion}
To estimate the parametric gain for test mass acoustic modes it is necessary to model all the acoustic modes in the frequency range from 5kHz to 150kHz for typical test mass dimensions. We used finite element modeling based on typical Advanced LIGO type test masses to model the acoustic mode structure of many thousands of acoustic modes combined with detailed modeling of the transverse optical mode structure of interferometer cavities~\cite{Evans2010}.

We modeled fused silica test masses similar to those used in Advanced LIGO and Advanced Virgo.  The detailed parameters for the fused silica test mass and optical cavity configuration can be found in reference ~\cite{Evans2010}. The key parameters are $P_{\text{cav}}=$830$kW$, acoustic $Q$ factors $(0.5\sim1)\times10^7$, coating loss limited. Based on this data we were able to predict the parametric gain of each test mass mode in an interferometer and determine the dependence of parametric gain on test mass mirror radius of curvature.  For comparison, we also included analysis of test masses similar to those proposed for use in KAGRA ~\cite{Somiya} (although as a cryogenic interferometer, the noise in KAGRA would be reduced).  Sapphire test masses have roughly 5 times lower acoustic mode density in the frequency range of interest, so that the number of relevant acoustic modes is $\sim1000$ in an interferometer with sapphire test masses, compared with $\sim5500$ for an interferometer with fused silica test masses.

\begin{figure}
\begin{center}
  \includegraphics[width=0.9\textwidth]{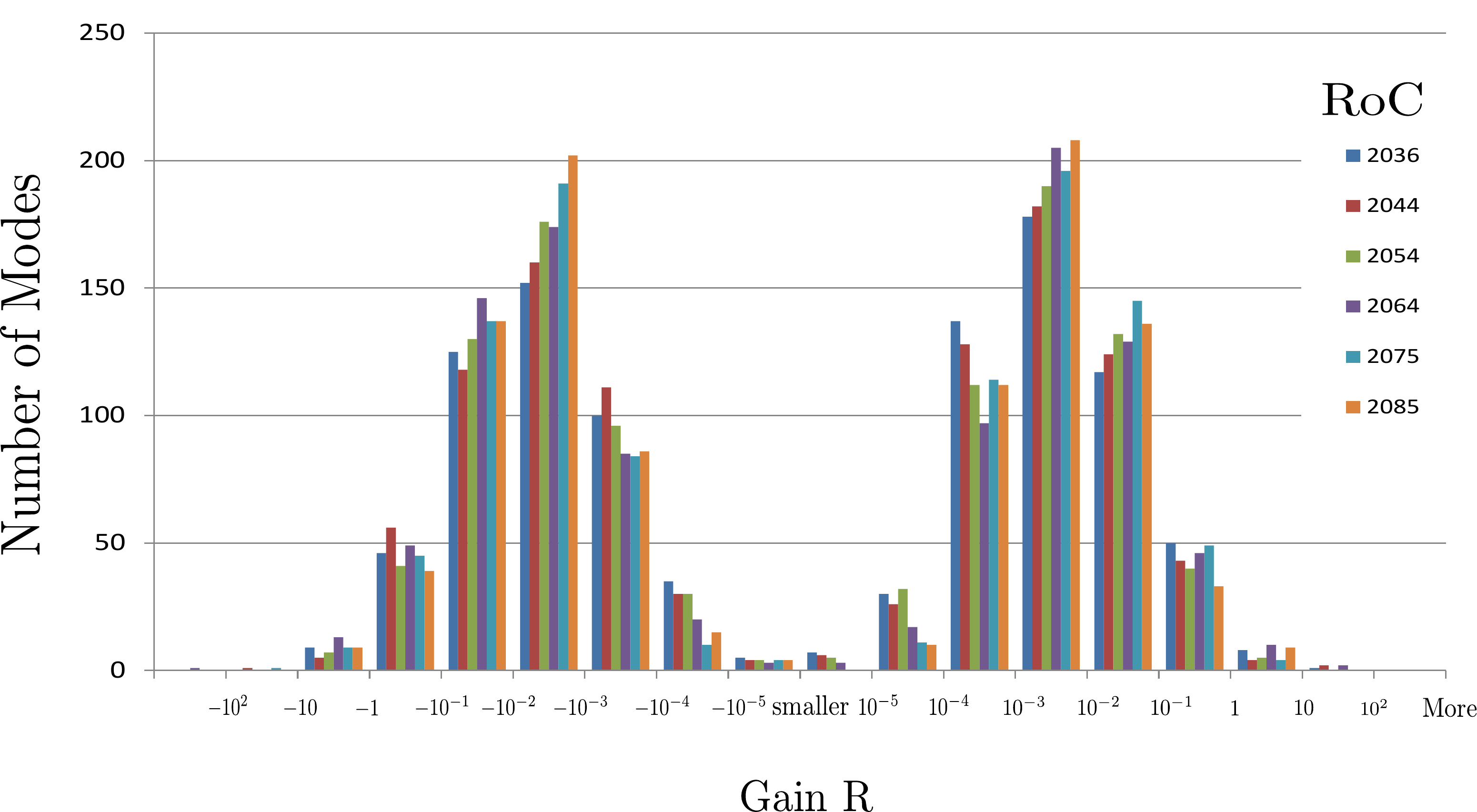}\\
  \caption{Figure 2 Histogram of the number of modes of one mirror with different parametric gain at different radius of curvature for typical advanced detector configuration (sapphire test mass).  Although the parametric gain depends strongly on mirror radius of curvature for particular acoustic modes, the overall statistical distribution of parametric gain is similar over large range of the Radius of Curvature (RoC).}\label{fig2}
\end{center}
\end{figure}

The radius of curvature (RoC) in a long interferometer varies with absorbed power, can be tuned by thermal compensation. Changes in RoC change the optical cavity mode structure and strongly tune the parametric gain of individual modes.  However, we find that the statistical distribution of the number of modes with different gain is similar over quite a large range of radii of curvature. An example is the analysis of the gain distribution of modes in sapphire test masses shown in Fig.~\ref{fig2}.
This figure means that a good estimate of three mode interaction noise can be obtained by summing over acoustic modes using a nominal value for radius of curvature. However the large difference in mode density means that the total noise is likely to be higher for fused silica.
\begin{figure}
\begin{center}
  \includegraphics[width=0.8\textwidth]{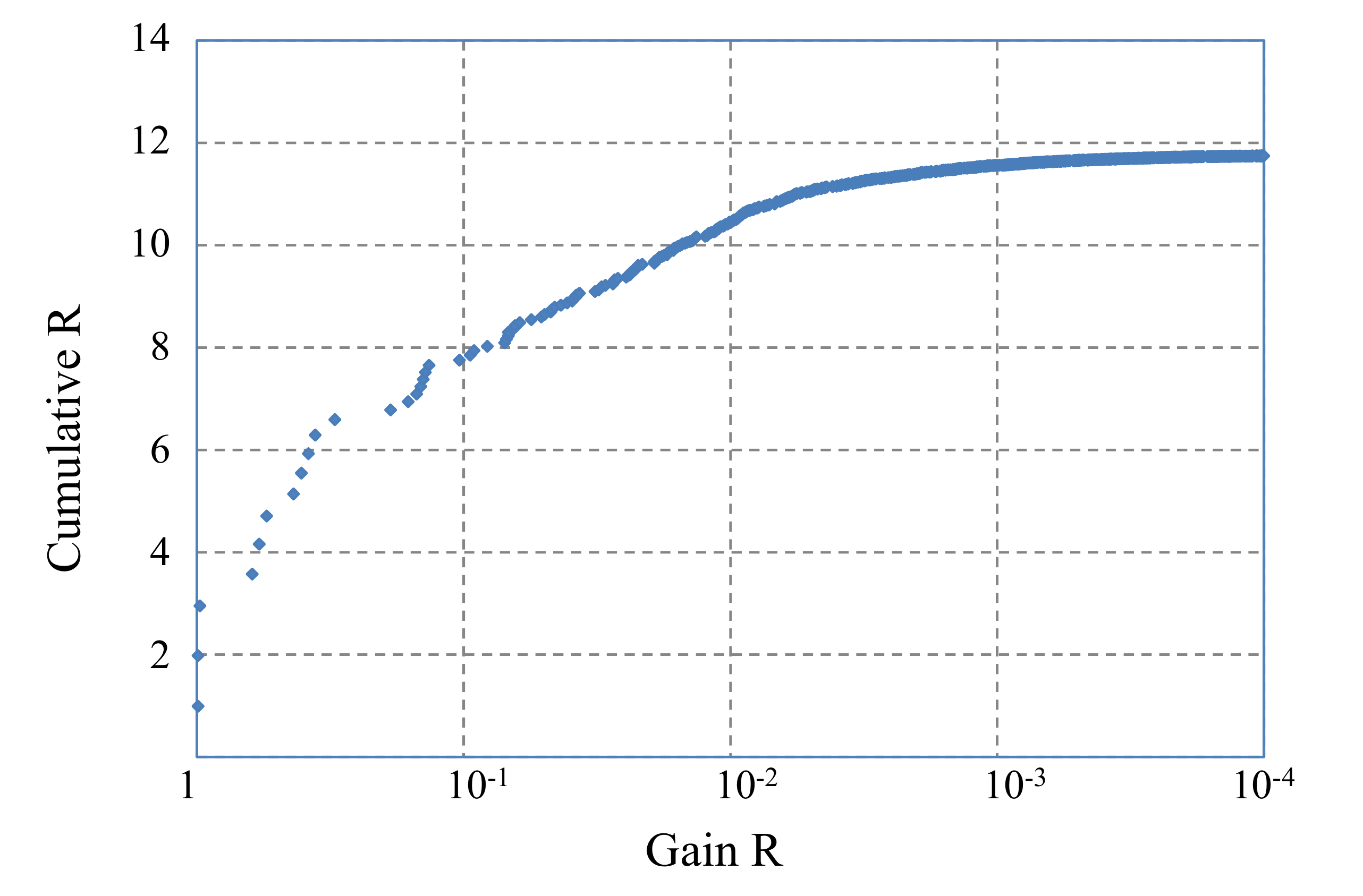}\\
  \caption{Cumulative R for positive gain of a fused silica test mass at RoC of 2191m.  It can be seen that modes with $1<R<0.1$ contribute $\sim60$ percent to the total $R$ while $R<1\times 10^{-3}$ has negligible contribution.}\label{fig3}
\end{center}
\end{figure}

Because the total noise depends on $R^2/(1-R)^3$, it is clear that
the noise will be dominated by any mode that has $R\rightarrow 1$.\
However, if $R<1$, the noise can be roughly estimated from the cumulative value of $R$ summed over all modes.
Fig.~\ref{fig3} gives model results showing the cumulative sum of $R$ for positive gain modes of a fused silica test mass at RoC of 2191m. This
model included one mode with $R\sim 50$ and several modes with $R>1$. Such
modes would normally be controlled using damping or feedback and hence would not contribute to the cumulative gain in an operating interferometer. The modes with gain between 1 and $10^{-2}$ contribute to total gain $\sim 20$, while modes with gain $10^{-2}\sim 10^{-3}$ contribute $R\sim 3$. Modes with $R<10^{-3}$ make a negligible contribution to the noise.

\begin{figure}
\begin{center}
\begin{tabular}{cc}
\includegraphics[width=0.55\textwidth]{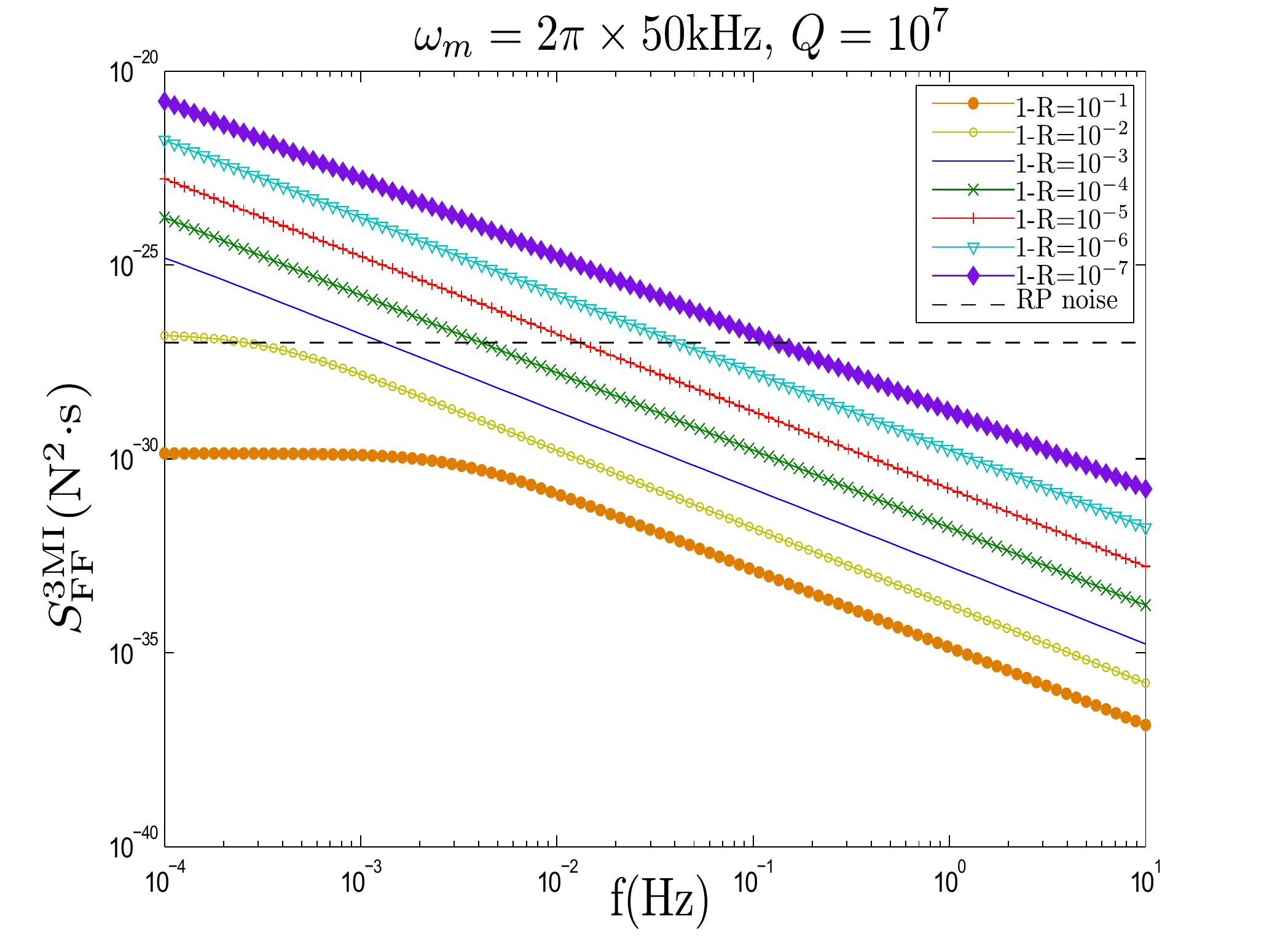}
&\includegraphics[width=0.51\textwidth]{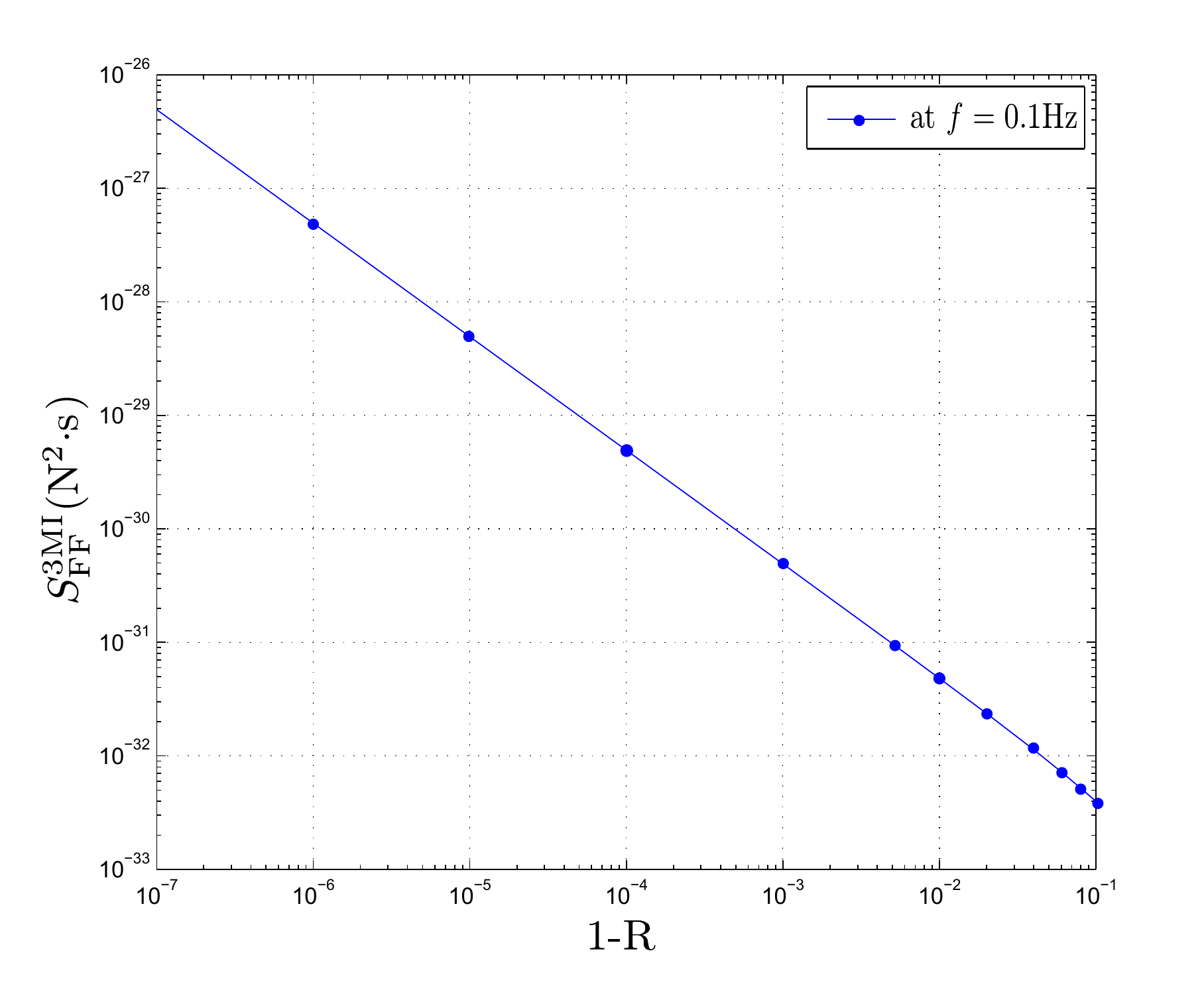}
\end{tabular}
\caption{The radiation pressure noise spectrum due to three-mode interactions for the advanced detector configuration considered in ~\cite{Gras2009} (a) Dependence of the noise force spectral density on parametric gain for a single mechanical mode. The dashed line is the radiation pressure noise for advanced detectors. (b) Force spectral density at $0.1$Hz as a function of $1-R$. Both models assume a mechanical mode frequency of 50kHz and $Q$-factor of $10^7$. Values of $1-R$ as small as $10^{-7}$  are considered because it is likely that drift in operating conditions of an interferometer could cause the gain to pass continuously from the stable regime $R<1$ to the unstable regime $R>1$.}
\label{Fig4}
\end{center}
\end{figure}

It is interesting to examine the effect of different R values on the radiation pressure force fluctuation for a single mechanical mode
(Eq.~\ref{S3MIfinal}).  Consider a mode of frequency 50kHz with a Q-factor of $1\times10^7$.  The radiation force fluctuation spectral density $S^{\text{3MI}}_{\text{FF}}$ is as shown in Fig.~\ref{Fig4}.  It can be seen that the fluctuations are at very low frequency due to the long relaxation time of the high mechanical $Q$-factor. The noise level increases strongly as $R$ approaches to unity. Normally $R$ would be expected to be controlled below unity to prevent parametric instability. However, if thermal drifts in the interferometer cause a particular mode to drift towards instability, then we would expect $R$ to increase asymptotically to 1. Fig.\ref{Fig4} shows the large increase of noise that would occur as $1-R$ falls to $10^{-7}$. This would not affect interferometer operation but is an interesting phenomenon in its own right.

It is shown in Fig.~\ref{Fig4} that the noise for one mode only approaches level of the quantum radiation pressure noise for $1-R<10^{-2}$. In the rare situation of parametric gain drifting towards 1, the three-mode interaction noise only exceeds the quantum radiation pressure noise spectrum much below the advanced LIGO frequency band (see Fig.~\ref{Fig4}).

\begin{figure}
\begin{center}
\includegraphics[width=0.8\textwidth]{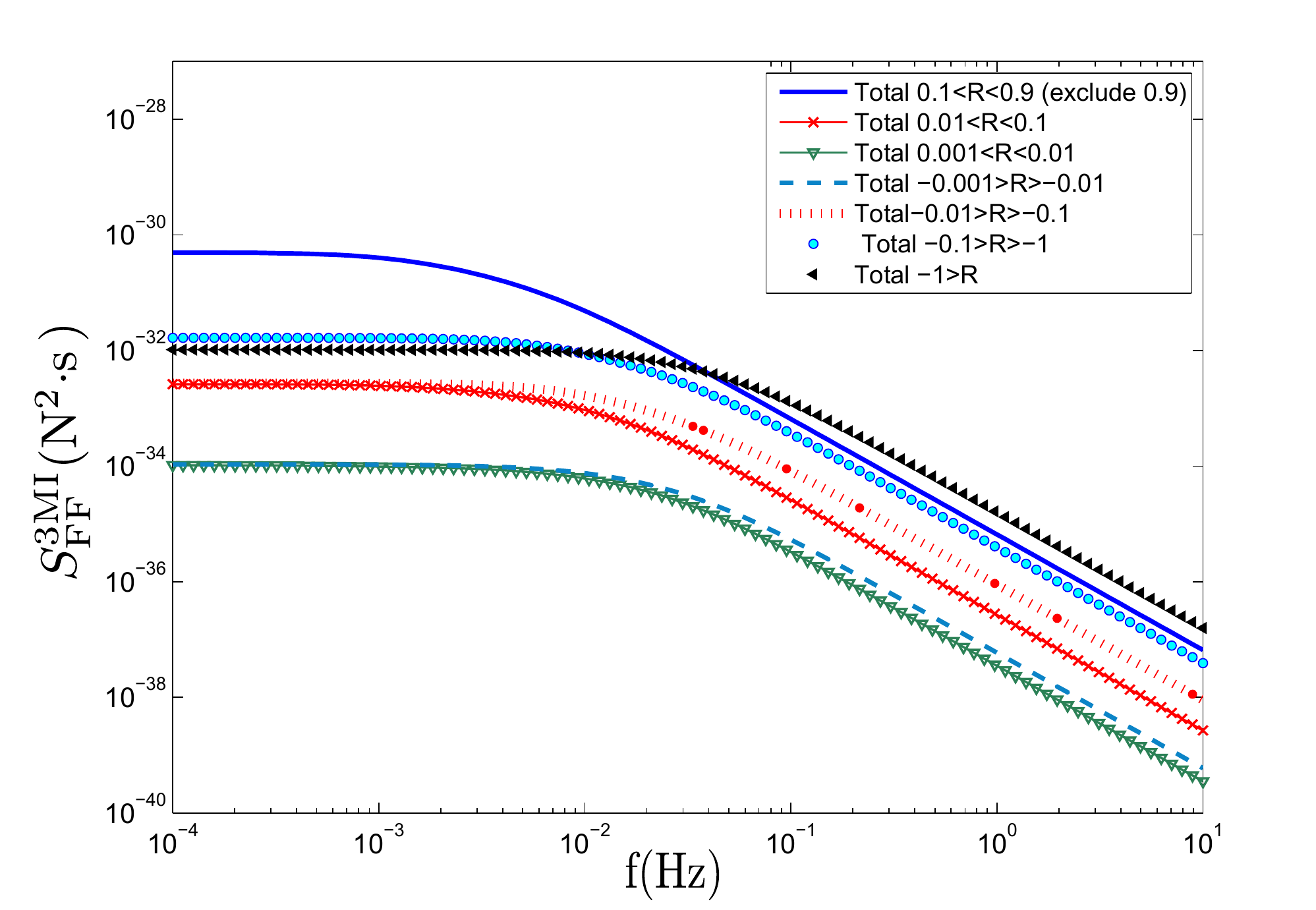}\\
\caption{Total contribution of modes with different R value to the noise. Note here the highest R value is $<0.9$, which correspond to the lowest curve in Fig.~\ref{Fig4}.  The result has assumed
that four test masses contribute to the total noise.}
\label{finalfig}
\end{center}
\end{figure}

Fig.~\ref{finalfig} shows the contribution to three mode interaction noise from the combined effects of all acoustic modes used in our model. The curves show the relative noise contribution from a) all modes with $0.9>R>0.1$, b) all modes with $0.1>R>0.01$, and c) all modes with $0.01>R>0.001$. It is clear that three mode interaction noise is likely to be dominated by the highest gain modes. The very large population of lower gain modes has a small contribution to the total noise.

For a fused silica test mass in advanced detector configuration, there could be about 10 modes with $R>1$.  We assume that all the unstable modes are controlled to $R<1$.  It is likely that a few modes could have parametric gain be very close to unity.   However it can be seen that even with $R\rightarrow 1$, the noise level due to three mode interactions is still low compared with radiation pressure noise in the frequency band of interest for advanced detectors.

Changing the test mass acoustic loss has a direct effect on the parametric gain distribution (Fig.~\ref{fig2}) as well as the spectral distribution.  If the $Q$-factor of ultrasonic acoustic modes were reduced as a whole, say by using an acoustic damper scheme (opposed to selective damping a few high parametric gain modes), then the total level of 3 mode interaction noise will drop. Similarly, increasing the power level increases the parametric gain, and hence increases the three mode interaction noise.

\section{Conclusion}
We have shown that three mode interactions give rise to a new source of intensity fluctuations in gravitational wave detectors. The noise is due to the effect of light scattering off test mass acoustic modes. Analysis of typical test masses for advanced gravitational wave detectors shows that the acoustic modes contribution to the noise spectrum peaks at frequencies well below the lower frequency limit of ground based gravitational wave detectors. The amplitude of three-mode interaction noise is substantially smaller than the quantum noise within the advanced detector frequency range. Thus the three-mode interaction noise will not limit the current generations of detectors but it should be considered when designing advanced quantum measurement schemes that can surpass the free mass standard quantum limit. It should also be considered when designing future detectors with larger test masses and sensitivity in the 1-5Hz range.

\section{Acknowledgements}
The authors want to thank Haixing Miao and Huan Yang for their helpful discussion. This work has been supported
by the Australian Research Council. Y. M  is also supported
by the Department of Education, Science and Training of Australia.

\appendix
\section{Two-points correlation function for $x^2$}
In the main text, we frequently use the two-points correlation function for $x^2$. In this appendix, we give a
analytic derivation of it.

As we know the $x(t)$ is given by:
\be
x(t)=\int^{t}_{-\infty}dt^{\prime}G(t-t^{\prime})\xi_{th}(t^{\prime}),
\ee
in which the  $\xi_{th}(t)$ is the thermal fluctuation force and
\be
G(t-t^{\prime})=\frac{1}{m\omega_m}\sin{\omega_m(t-t^{\prime})}
e^{-\frac{\gamma_m}{2}(t-t^{\prime})}
\ee
is the Green's function for a damped mechanical oscillator.
Substitute it into $\langle x^2(t)x^2(t+\tau)\rangle$, we have:
\be
\begin{split}
\langle x^2(t)x^2(t+\tau)\rangle
=\frac{D^2}{m^4\omega_m^4}\int^{t}_{-\infty}\int^{t+\tau}_{-\infty}
dt^{\prime}dt^{\prime\prime}dt^{\prime}_ldt^{\prime\prime}_l
e^{-\frac{\gamma_m}{2}(t-t^{\prime})}e^{-\frac{\gamma_m}{2}(t-t^{\prime\prime})}
e^{-\frac{\gamma_m}{2}(t+\tau-t^{\prime}_l)}e^{-\frac{\gamma_m}{2}(t+\tau-t^{\prime\prime}_l)}\\
\left[\sin{\omega_m(t-t^{\prime})}\sin{\omega_m(t-t^{\prime\prime})}\sin{\omega_m(t+\tau-t^{\prime}_l)}
\sin{\omega_m(t+\tau-t^{\prime\prime}_l)}\right]
\langle\xi(t^{\prime})\xi(t^{\prime\prime})\xi(t^{\prime}_l)\xi(t^{\prime\prime}_l)\rangle
\end{split}
\ee
in which $D=4m\gamma_mk_BT$.

Assuming that thermal force is a Gaussian process, therefore it is straightforward to see that:
\be
\langle\xi(t^{\prime})\xi(t^{\prime\prime})\xi(t^{\prime}_l)\xi(t^{\prime\prime}_l)\rangle=\delta(t^{\prime}-t^{\prime\prime})
\delta(t^{\prime}_l-t^{\prime\prime}_l)+\delta(t^{\prime}-t^{\prime}_l)\delta(t^{\prime\prime}-t^{\prime\prime}_l)+
\delta(t^{\prime}-t^{\prime\prime}_l)\delta(t^{\prime\prime}-t^{\prime}_l)
\ee

Substitute (A.4) into (A.3) after a complicated but straightforward calculation, we have:
\be
\begin{split}
\langle x^2(t)x^2(t+\tau)\rangle=\frac{D^2}{m^4\omega_m^4}\left[\frac{4\omega_m^4}{(\gamma_m^3+4\gamma_m\omega_m^2)^2}
+\frac{4\omega_m^4}{(\gamma_m^3+4\gamma_m\omega_m^2)^2}e^{-\gamma_m\tau}\cos^2{\omega_m\tau}
\right]\\
+2\frac{D^2}{m^4\omega_m^4}\left[\frac{2\omega_m^2}{(\gamma_m^2+4\omega_m^2)^2}e^{-\gamma_m\tau}\sin^2{\omega_m\tau}
+\frac{4\omega_m^3}{\gamma_m(\gamma_m^2+4\omega_m^2)^2}e^{-\gamma_m\tau}\sin{2\omega_m\tau}\right]
\end{split}
\ee

Since our $Q$-factor is high, therefore we can reexpress the above formula as expansion of $Q$:
\be
\begin{split}
\langle x^2(t)x^2(t+\tau)\rangle=\frac{D^2}{m^4\omega_m^4}\left[\frac{4\omega_m^2Q^2}{(\gamma_m^2+4\omega_m^2)^2}
+\frac{4\omega_m^2Q^2}{(\gamma_m^2+4\omega_m^2)^2}e^{-\gamma_m\tau}\cos^2{\omega_m\tau}
\right]\\
+2\frac{D^2}{m^4\omega_m^4}\left[\frac{2\omega_m^2}{(\gamma_m^2+4\omega_m^2)^2}e^{-\gamma_m\tau}\sin^2{\omega_m\tau}
+\frac{4\omega_m^2Q}{(\gamma_m^2+4\omega_m^2)^2}e^{-\gamma_m\tau}\sin{2\omega_m\tau}\right]
\end{split}
\ee

Apparently, the $Q^2-$terms are dominate, thereby we have:
\be
\langle x^2(t)x^2(t+\tau)\rangle=\frac{D^2}{m^4\omega_m^4}\frac{4\omega_m^2Q^2}{(\gamma_m^2+4\omega_m^2)^2}\left[
1+2e^{-\gamma_m\tau}\cos^2{\omega_m\tau}
\right]
\ee

Make further approximation that $\omega_m\gg\gamma_m$, we have:
\be
\langle x^2(t)x^2(t+\tau)\rangle=\frac{D^2}{m^4\omega_m^4}\frac{4\omega_m^2Q^2}{16\omega_m^4}\left[
1+2e^{-\gamma_m\tau}\cos^2{\omega_m\tau}.
\right]
\ee

Substituting $D=4m\gamma_mk_BT$, we have:
\be
\langle x^2(t)x^2(t+\tau)\rangle=\frac{4k_B^2T^2}{m^2\omega_m^4}\left[
1+2e^{-\gamma_m\tau}\cos^2{\omega_m\tau}
\right]
\ee

The high frequency $2\omega_m$ part can be neglected, therefore
the mechanical displacement four-point correlation function can be expressed as::
\be
\langle x^2(t)x^2(t+\tau)\rangle=\frac{4k_B^2T^2}{m^2\omega_m^4}\left[
1+e^{-\gamma_m\tau}\right]
\ee

Here we should notice that this result tells us that
the stochastic process $x(t)$ is a \emph{Gaussian process} under the
approximation we made above. The reason is in this case $x(t)$
obeys the \emph{Wick theorem}:
\be
\langle x(t)x(t^{\prime})x(t^{\prime\prime})x(0)\rangle
=\langle x(t)x(t^{\prime})\rangle\langle x(t^{\prime\prime})x(0)\rangle
+\langle x(t)x(t^{\prime\prime})\rangle\langle x(t^{\prime})x(0)\rangle+\langle x(t)x(0)\rangle\langle x(t^{\prime})x(t^{\prime\prime})\rangle
\ee
Since the two point function of mechanical displacement for a
high-Q oscillator in a thermal bath in the time domain can be written as:
\be
\langle x(t)x(t+\tau)\rangle=\frac{2k_BT}{m\omega_m^2}e^{-\frac{\gamma_m}{2}\tau}\cos{\omega_m\tau}
\ee
Then (A.11) can be written as:
\be
\begin{split}
\langle x^2(t)x^2(t^{\prime})\rangle
=\langle x^2(t)\rangle\langle x^2(t+\tau)\rangle+
2\langle x(t)x(t+\tau)\rangle^2\\
=\frac{4k_B^2T^2}{m^2\omega_m^4}(1+e^{-\gamma_m\tau}2\cos^2{\omega_m\tau})
=\frac{4k_B^2T^2}{m^2\omega_m^4}\left[1+e^{-\gamma_m\tau}(1+\cos(2\omega_m\tau))\right]
\end{split}
\ee
Neglect the high frequency $2\omega_m$ term, we recover the result (A.10).
\end{document}